\documentclass[number,citesort,dvips]{arxstspdf}
\usepackage{sgmlsup}
\usepackage{graphicx,flushend,stfloats}

\makeatletter
\setattribute{abstract}    {skip} {15\p@}
\setattribute{keyword}     {skip} {1\p@}
\setattribute{frontmatter} {skip} {0\p@ plus 3\p@ minus 3\p@}
\setattribute{frontmatter} {cmd}  {}
\setattribute{abstract}   {width}  {28pc}
\setattribute{keyword}    {width}  {28pc}
\makeatother

\volume{25}
\issue{2}
\pubyear{2010}
\firstpage{258}
\lastpage{273}
\doi{10.1214/08-STS272}

\begin{document}
\begin{frontmatter}

\title{A Conversation with Martin Bradbury Wilk}
\runtitle{A Conversation with Martin B. Wilk}

\begin{aug}
\author[a]{\fnms{Christian} \snm{Genest}\ead[label=e1]{Christian.Genest@mat.ulaval.ca}\corref{}} \and
\author[b]{\fnms{Gordon} \snm{Brackstone}\ead[label=e2]{Gordon.Brackstone@sympatico.ca}}
\runauthor{C. Genest and G. Brackstone}

\affiliation{Universit\'{e} Laval, Statistics Canada}

\vspace*{-7pt}
\address[a]{Christian Genest is Professor of Statistics, D\'{e}partement de
math\'{e}matiques et de statistique, Universit\'{e} Laval, 1045, avenue
de la M\'{e}decine, Qu\'{e}bec, Qu\'{e}bec, Canada G1V 0A6
\printead{e1}.}
\address[b]{Gordon Brackstone is retired; he was formerly the Assistant Chief
Statistician, Informatics and Methodology Field, Statistics Canada
\printead{e2}.}

\end{aug}

\begin{abstract}
\fontsize{10.3}{12.3}\selectfont
Martin Bradbury Wilk was born on December 18, 1922, in
Montr\'{e}al, Qu\'{e}bec, Canada. He completed a B.Eng. degree in
Chemical Engineering in 1945 at McGill University and worked as a
Research Engineer on the Atomic Energy Project for the National Research
Council of Canada from 1945 to 1950. He then went to Iowa State College,
where he completed a M.Sc. and a Ph.D. degree in Statistics in 1953 and
1955, respectively. After a one-year post-doc with John Tukey, he became
Assistant Director of the Statistical Techniques Research Group at
Princeton University in 1956--1957, and then served as Professor and
Director of Research in Statistics at Rutgers University from 1959 to
1963. In parallel, he also had a 14-year career at Bell Laboratories,
Murray Hill, New Jersey. From 1956 to 1969, he was in turn Member of
Technical Staff, Head of the Statistical Models and Methods Research
Department, and Statistical Director in Management Sciences Research. He
wrote a number of influential papers in statistical methodology during
that period, notably testing procedures for normality (the Shapiro--Wilk
statistic) and probability plotting techniques for multivariate data. In
1970, Martin moved into higher management levels of the American
Telephone and Telegraph (AT\&T) Company. He occupied various positions
culminating as Assistant Vice-President and Director of Corporate
Planning. In 1980, he returned to Canada and became the first
professional statistician to serve as Chief Statistician. His
accomplishments at Statistics Canada were numerous and contributed to a
resurgence of the institution's international standing. He played a
crucial role in the reinstatement of the Cabinet-cancelled 1986 Census.
He remained active after his retirement, serving as a Senior Advisor to
the Privy Council Office as well as on several national commissions. In
addition, he chaired the Canadian National Task Forces on Tourism Data
and on Health Information. Martin is a former President of the
Statistical Society of Canada (SSC) and Vice-President of the American
Statistical Association (ASA). He is an elected member of the
International Statistical Institute and an honorary member of the SSC.
He has received many honors, including the George Snedecor Prize, the
Jack Youden Prize, the F.G. Brander Memorial Award, the SSC Gold Medal,
and a Distinguished Alumni Achievement Citation from Iowa State
University. He is a fellow of the Institute of Mathematical Statistics,
the American Statistical Association, the Royal Statistical Society, the
American Association for the Advancement of Science, and the New York
Academy of Science. He was made an Officer of the Order of Canada in
1999 for his ``insightful guidance on important matters related to our
country's national statistical system.''

The following conversation took place at Martin Wilk's home in Salem,
Oregon, October 6--7, 2005. 
\end{abstract}


\end{frontmatter}

\section*{Introduction}

\textbf{Christian}: Martin, tell us something about your youth.

\begin{figure}

\includegraphics{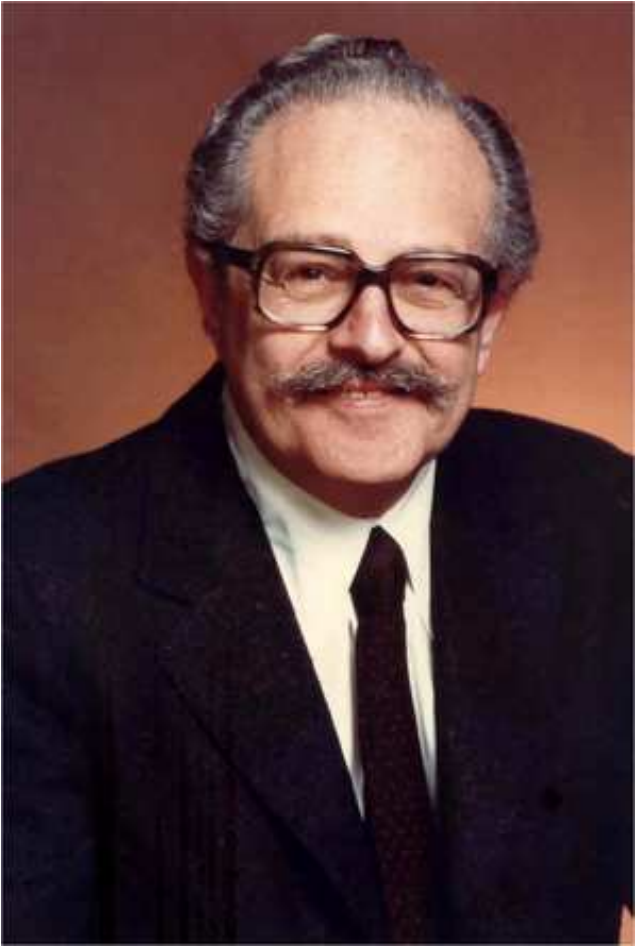}

\caption{Official photograph of Martin Bradbury Wilk, circa 1985.}
\end{figure}

\textbf{Martin}: I was born in Montr\'{e}al in December 1922. I was the
youngest in a family of three children. I have two sisters who are still
alive today. We lived in a part of town where English was dominant, and
I should immediately confess to you, Christian, that although I had many
years of training in French, I ended up with virtually no capacity in
that language. I think I'm just one of those people, unfortunately, who
inherited not much capability for a second tongue. But I dare say, it
wasn't much of a handicap in Montr\'{e}al, at least in those days.

I got my secondary education at the now defunct Strathcona Academy in
Outremont [now part of Mont-r\'{e}al]. I had no burning academic interest
at the time. Mathematics and geometry were the topics I thought were
easiest. In Grade 11, which was the final year, my worst grades were in
French, oral and written. Other than that, all topics seemed pretty
straightforward, and I guess I learned some bad working habits as a
result.

\section*{Engineering at M{\lowercase{c}}Gill}

\textbf{Christian}: After graduation from Strathcona in 1940, you
studied engineering at McGill University, in Montr\'{e}al. What guided
your choice?

\textbf{Martin}: When I was young, I was generally interested in the
technical side of things, and McGill seemed like the place to go. It was
one of the largest and most prestigious universities in Canada. It had
a~big reputation, especially in medicine. There was a year of transition
with English literature, history, mathematics, and so on. Then you could
go for a four-year program in the Engineering Faculty.

I found that first year really easy, but once I got into engineering
that turned out to be quite a different story. The first two years of
the engineering program were common to all. We were about 130 people at
the start, and whoever gave the welcoming speech made a remark to the
effect that half of us would be gone by the end of the first year. And
he was right.

The first year in engineering was really hard for me. My bad habits were
a disservice to me, and I ended up being second to last in the list of
people who made the cut! Luckily, my performance began to improve at
that point, and I was able to complete the program.

\textbf{Gordon}: What got you interested in chemical engineering
specifically?

\textbf{Martin}: I enjoyed chemistry in high school, although we never
had a chance to use a laboratory, because that wasn't part of the
understanding at the time. Once I got to McGill, I found out that I was
very good hands-on. Also, there was an arrangement at that time that you
had to be employed for the summer and write a term paper with regard to
your experience. And as it turned out, all my summer jobs had to do with
chemistry.

\section*{Summer Jobs}

\textbf{Christian}: Where did you work as a student?

\textbf{Martin}: At Howard Smith's paper mill the first summer. They
operated out of Cornwall, Ontario. Then a year later, I worked for
Distillers Corporation in Montr\'{e}al. This is a company that Samuel
Bronfman had founded; it produced commercial alcohol. It had enjoyed
substantial growth in the 1920s, due to Prohibition in the States. But
during the war, the company was busy making pure alcohol for the army.

In my final summer before graduation, I worked for the Montr\'{e}al Coke
and Manufacturing Company. It was the most interesting of my summer
jobs. It was a dry distillation of coal to produce coke. It was done in
drying ovens whose temperature had to be monitored very closely, 24
hours a day. There were data to be collected and I was involved in that.
I remember going up the chimney, wearing a sweater to keep me from being
burned, and gathering data.

\begin{figure}

\includegraphics{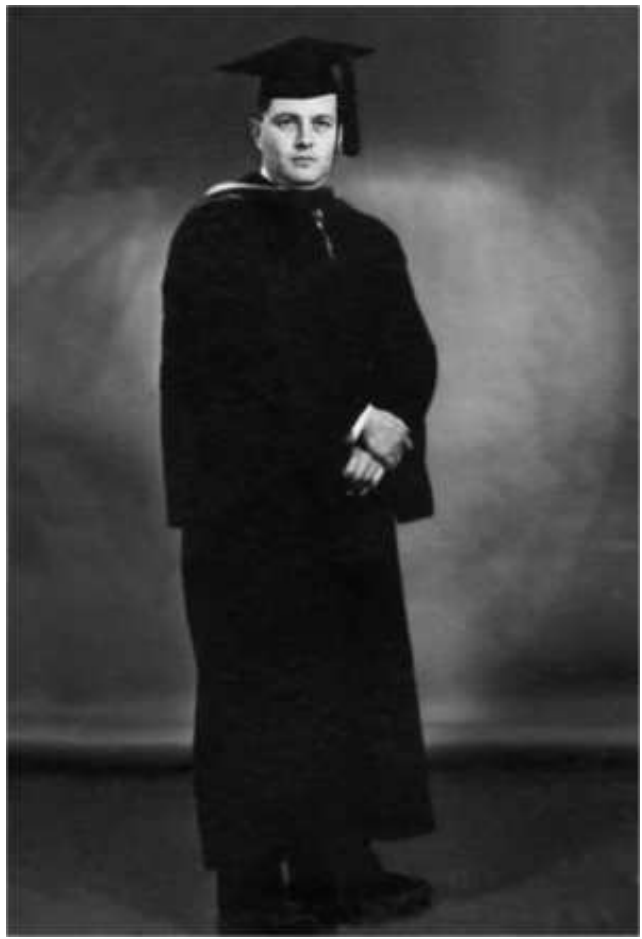}

\caption{Martin's graduation photograph, McGill University,
Montr\'{e}al, 1945.}
\end{figure}

\textbf{Gordon}: Did the war affect you in any way?

\textbf{Martin}: Not really. There was no conscription in Canada in the
early 1940s. However, we did have to come to school in uniform twice a
week, and spend some time at so-called training. I also learned how to
identify a flying aircraft and to read Morse code along the way. But
more significantly, perhaps, is that when I graduated in 1945, the war
was still on. So I was offered a choice between doing a Ph.D. in
chemistry at McGill or joining Canada's National Research Council (NRC).

\textbf{Christian}: You make it sound as though you had no other option.
Was that really the case?

\textbf{Martin}: Well, I mean I was told ``That's where you're going.''
And at the time it was part of the understanding that you do as you're
told. I mean, the alternative to that was you go in the army.

\textbf{Gordon}: And why didn't you go for the Ph.D.?

\textbf{Martin}: It paid well: \$75 a month, if I recall. This was a
substantial amount of money at the time. But I didn't have the patience.
You see, I don't think I'm a natural student. This is something I
learned subsequently. And I found that listening to people telling me
about things that they have done, or about things that other people have
done, was not part of my nature.

\section*{Chalk River Laboratories}

\textbf{Christian}: What was your position at NRC?

\textbf{Martin}: When I was hired in June, 1945, I moved to Ottawa and
spent about six months there. Then I joined the Chalk River
Laboratories, located in Deep River, in the upper Ottawa valley. The
first nuclear reactor outside of the United States had just gone
operational up there. There were about 250 scientists on the site.
Nuclear technology was the focus of our activity.

There were many people from abroad. One of the most remarkable figures
was Bruno Pontecorvo, an Italian physicist who had worked with Enrico
Fermi. I got to know him because each of us separately had a bad habit
of missing the bus that drove us to work in the morning. Pontecorvo was
probably a genius, and I am not using that term loosely. It came to me
as a great surprise when he defected to Russia in 1951, in the middle of
the Cold War.

\textbf{Gordon}: And what were your responsibilities?

\textbf{Martin}: I was mostly in charge of testing an air cooling system
for the rods that were used in the experimental heavy-water pile. I
gathered a lot of data in this context and under the best of
circumstances, there was a great deal of variability associated with it.
Radioactive behavior is by-and-large unpredictable, except on average.
So I began developing a few techniques of my own to handle such data.

\section*{Statistics Training in Iowa}

\textbf{Christian}: Does this explain why you chose to leave Chalk River
in 1950 to do graduate work in statistics at Iowa State College?

\begin{figure}

\includegraphics{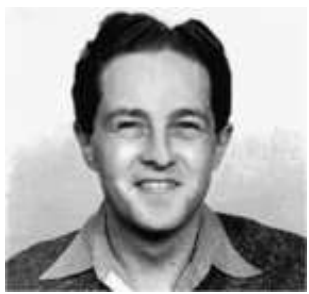}

\caption{Martin's Photo ID when he applied for graduate school in
statistics at Iowa State College.}
\end{figure}

\begin{figure*}[b]

\includegraphics{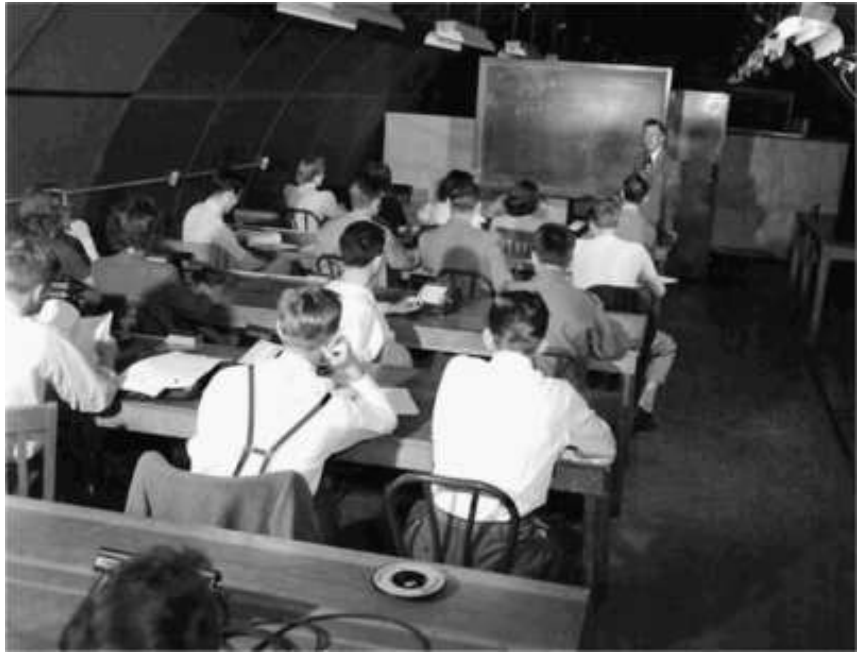}

\caption{Bernie Ostle teaching statistics students at Iowa State
College, circa 1950.}
\end{figure*}

\textbf{Martin}: Well, not quite. Let me explain. My time in Chalk River
was very pleasant and productive. I even had a paper published there
\cite{1}. But after five years, I~had a feeling that if I stayed longer, I'd
probably be there for the rest of my life. And I should mention
something here, which at the time I had no knowledge of, but it's really
turned out that throughout my career, I had a different job every five
years. Not that I ever manipulated or arranged for it, but it just so
happened.

Anyway, it so happened that some friends of mine had gone to Ames, Iowa.
They were microbiologists and enjoyed it there. So they encouraged me
to join them and even went as far as negotiating a contract for me with
a chap called R. G. Tischer, who operated in food technology. He offered
me a position as a research assistant and the salary was enough to make
me feel that I could go.

\textbf{Gordon}: So you went just like that?

\textbf{Martin}: To tell you the truth, it was sort of a mindless
decision, like many I took in my life. I didn't really think hard about
it and don't recall that I needed to be particularly brave. And
actually, once I got there, I found out that Professor Tischer's
research program was much too prosaic in my outlook. I do recall
spending countless hours in the lab, carrying out experiments for him,
and reading instruments, etc. But I was quickly bored.

Now Professor Tischer knew enough about his own subject to recognize
that there was variability in the observations he was dealing with, and
given my prior exposure to data in Chalk River, I could provide some
assistance with this. Eventually, he encouraged me to sign up for a
statistics course, which I did, and I was interested to discover that
there was a theory behind much of what I had basically developed on my
own, in some form.

\begin{figure*}[b]

\includegraphics{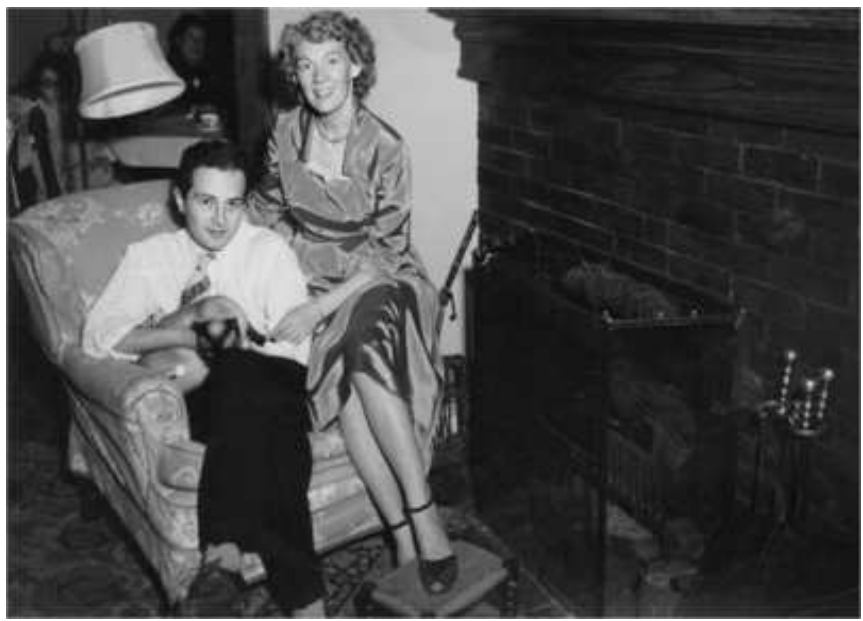}

\caption{Martin and his first wife, Thora Sugrue, at their home in
Neshanic, New Jersey, circa 1957.}
\end{figure*}

\textbf{Christian}: Do you remember who taught you that course?

\textbf{Martin}: It was Bernie Ostle. He had me meet with various people
in the Statistics Department, and he encouraged me to solidify my
mathematical encounters. I~began doing that, and it all came very
naturally. I~mean, I really never had in mind that I was going to leave
chemical engineering, but before the first year in Ames was over, I
decided that it wasn't right for me to continue taking money from
Tischer. I decided to pursue the mathematics I was engaged in and work
toward a Master's degree.

\textbf{Gordon}: Who was your advisor?

\textbf{Martin}: It was Oscar Kempthorne. He supervised my Master's
thesis, which I completed in 1953, and then he persuaded me to stay on
for the Ph.D., which I got in 1955. My thesis was in the area of linear
models and the analysis of variance for randomized block designs. It led
to several publications \cite{2,3,4,5,6}, some of them joint with Oscar.

\textbf{Christian}: What was it like to work with Professor Kempthorne?

\textbf{Martin}: He was a man of substantial intellect and he had very
strong feelings about a variety of matters, technical, and nontechnical.
I shared an office with him throughout most of my studies in Iowa, and
we grew to be very good friends. He tended to be quite outspoken and to
express his opinions emphatically. This may explain in part why the
department was polarized, or at least separated between those people
around Herman Hartley, who were involved in survey sampling, and those
like Oscar Kempthorne and I, who were more concerned with analysis of
variance and experimental design.

\textbf{Gordon}: Who else was there at the time?

\textbf{Martin}: People like Bernie Ostle, John Gurland, Ted Bancroft,
and so on.

\textbf{Christian}: And you joined the group after you completed your
thesis, right?

\textbf{Martin}: No, actually I was established as an Assistant
Professor there before I completed my Ph.D. But as it turned out, Oscar
wanted me to gain additional experience before settling in at Iowa
State. So he got in touch with John Tukey, who was working at Princeton
and for Bell Laboratories, and he made arrangements for me to go there
for a year as a post-doctoral fellow.

I was actually quite pleased with this offer, because I had already met
briefly with Tukey at the meeting of the American Statistical
Association in Montr\'{e}al, in September 1954. I presented a joint
paper with Oscar Kempthorne at that meeting, and I encountered John
Tukey at that time. He already had a reputation as the ultimate wise
man, and my impression was one of awe. He struck me as a very
intelligent fellow, but it was only later that I found out how smart he
really was.

\section*{Post-doc at Princeton University}

\textbf{Christian}: Tell us about your post-doc year at Princeton
University then.

\textbf{Martin}: I arrived at Princeton in the spring of 1955. Sam Wilks
was about to leave for Texas and I inherited his office.

\textbf{Christian}: Wilks' office became Wilk's office, eh?

\textbf{Martin}: That was a good start! David Cox was there at the time;
he and I worked under Tukey's leadership that summer. Tukey had a grand
plan and lots of problems he wanted us to work on. He was the smartest
man I've ever associated with up close. His speed of thinking was quite
incredible, and his ability in mathematics was overwhelming. He was
always so far ahead that he could lecture me on what I was trying to do
far beyond what I could ever manage to accomplish. The~whole process was
very depressing and irritating to me. David Cox seemed to cope better; I
guess he had an interest and a capacity that I didn't have. But until
David left in early fall, the two of us would often commiserate on the
fact that Tukey was as smart as he actually was.

Anyway, my propensity is not to want to learn things unless I have a
motivation, so I thought my year in Tukey's environment at Princeton was
miserable. Maybe I didn't do as badly as I felt at the time, but in any
event, I was in pretty bad shape by the end of the year. At the same
time, my home life was hectic too. While doing my Master's thesis in
Ames, I had gotten married to Thora Sugrue and by that time we had four
young children. That drew a lot of energy!

\section*{Research at Bell Labs}

\textbf{Christian}: So how come you didn't go back to Iowa at the end
of the year?

\textbf{Martin}: Well, in the spring of 1956, John Tukey, who seemed to
know more about me than I knew about myself, said how would you like to
spend the summer at Bell Labs? I hadn't a clue what Bell Labs was, but
it sounded intriguing. So, at John's instigation, I went there and gave
a talk in front of the statistics group, which was under the direction
of Milton Terry at the time. I must have made a good impression because
at the end, they made me an offer and after some cogitation, I accepted.

\textbf{Christian}: Did they offer you a regular contract right then?

\textbf{Martin}: No, it was just a post-doc at first. But then John
Tukey arranged for me to work two days a week at Bell Labs and the rest
as an Assistant Director for the Statistical Techniques Research Group
at Princeton. So I got in touch with Oscar Kempthorne and Ted Bancroft,
who was the Department Chair, to let them know that I would not return
after all.

\textbf{Gordon}: What prompted your decision?

\textbf{Martin}: The first summer I spent at Bell Labs was a very
exciting time. To begin with, I encountered computer technology in the
guise of an IBM 650, which was an incredibly fast machine at the time.
Also, the friendly atmosphere and the freedom you enjoyed as a
researcher working for Bell Labs were exceptional. The staff was about
20,000 overall, and maybe 10\% of these were scientists doing research
in physics, chemistry, materials engineering, mathematics and
statistics, etc. The Research Department had a distinctive character,
and I was able to get involved in many projects as a statistical
consultant, and to some extent, as a chemical engineer, too. A couple of
publications ensued \cite{7,8}.

\textbf{Christian}: And what were your responsibilities at Princeton?

\textbf{Martin}: I was involved in research there too. But I~still
found that association to be disappointing.

\textbf{Gordon}: And why was that?

\textbf{Martin}: Chiefly because of my inability to support the work of
George Box, who had just been appointed Director of the Statistical
Techniques Research Group at Princeton. He was a very kind person, and
he had a quick and relatively creative mind, but our styles just didn't
match. He tended to be very focused in his research, and he was good at
it too, but I guess because of my background and through my contacts
with people at Bell Labs, I had a much broader view of things. Besides,
I had pretty well fallen in love with Bell Labs at that time. So I quit
Princeton after only a few months, but through Bell Labs, my association
with John Tukey lasted a long time.

\textbf{Christian}: Did your responsibilities at Princeton include any
teaching?

\textbf{Martin}: I had done a little bit of teaching in Iowa, but I did
none at Princeton. Where I did teach was at Rutgers, where I was a
professor of statistics from 1959 to 1963.

\section*{Professor at Rutgers}

\textbf{Gordon}: How did that come about?

\textbf{Martin}: I was approached in 1958 by Ellis Ott, who was heading
the Department of Statistics there. He wanted to establish a Ph.D.
program. He persuaded me to come down and teach an evening class in
mathematical statistics. I agreed, and then in January 1959 he came up
with the notion that I could be a full professor there and have the
responsibility for research. I~agreed, on the provision that I could
still do consulting and other work with Bell Labs, one day a week.

The arrangement appealed to me on several accounts. First, Rutgers paid
well and I needed that kind of money, given that I had a large family.
Second, I~thought maybe I'd like to work in a university environment and
lecture, as opposed to listening to other people. But I didn't want to
give up on Bell Labs completely. The colleagues at Bell Labs turned out
to be very encouraging in this regard.

Finally, an additional reason why Rutgers appealed to me was that Marion
Johnson, who was Dean of the Graduate School at the time, was close to
retirement and the indication was that I would be a good candidate for
replacement.

\textbf{Christian}: And did your expectations materialize?

\textbf{Martin}: I was on faculty at Rutgers four years in total, and I
certainly did my best to serve their interests. But to tell the truth, I
had a miserable time throughout that period, in that my wife Thora came
down with cancer. This episode started almost immediately after I joined
the faculty at Rutgers. I was at home almost round the clock, caring for
her and looking after the children. In the end, we lost the battle
against cancer and she died on April 15, 1965. Needless to say, I never
applied for the Dean's position at Rutgers.

Somehow, despite the hardships, I managed to be reasonably productive
throughout that period. No\break doubt, this was due in part to the great
collaborators I had. I am thinking especially of Ram Gnanadesikan and
Sam Shapiro. Another more prosaic factor is that throughout my entire
life as an adult, I never slept more than about four hours a night. Of
course, I would also ``cat nap'' on occasion during the day. In fact, I
did it even while I was Chief Statistician of Canada, but I'm getting
ahead of myself now.

\section*{Contributions to Statistical Methodology}

\textbf{Christian}: You made a large number of contributions to
statistical methodology in the 1960s. How did that develop?

\textbf{Martin}: Much of the research I did in that period was dictated
or inspired by questions of consultancy at Bell Labs. If you look at my
publications from 1960 to 1970 \cite{8,9,10,11,12,13,14,15,16,17,18,19,20,21,22,23,24,25,26,27,28,29,30,31,32,33}, you'll see that a prime concern
of mine was diagnostic procedures for classical distributions, e.g., the
normal or the exponential and the gamma. A fair portion of this work was
carried out with Ram Gnanadesikan, who was a colleague at Bell Labs, and
Samuel Shapiro.

Ram and I formed a highly compatible team. I think we wrote 12 papers
together. One thing about Ram is that because of his Indian origins, he
had a cultural bias about seniority and although each of us would do our
fair share of our joint work, he insisted that my name be listed first;
this continued for many years \cite{10,12,13,16,17,20,24,28}. As for
Shapiro, he was a student in one of my classes at Rutgers. He was a good
guy, and I agreed to supervise his dissertation. Our association
continued for a while \cite{21,26,27,29,34}.

At one point in time, it became clear to me that the behavior of order
statistics, in some sense, would have to reflect the nature of a
distribution. Ram, Sam, and I used the properties of these statistics to
design QQ-plots and goodness-of-fit tests.

\textbf{Christian}: The 1965 \textit{Biometrika} paper \cite{21} that
introduces the Shapiro--Wilk test statistic is certainly a classic.

\textbf{Martin}: I am obviously pleased with that, but to be truthful I
don't regard this work as such a great accomplishment. As I am fond of
saying, significance tests are things to do while one is trying to think
of something sensible to do. This being said, while it is true that the
idea behind the test was mine, Sam carried it considerably further with
power comparisons and approximations to the null distribution of the $W$
statistic, as he kept calling it.

\textbf{Christian}: This was truly seminal work. It is just a bit
unfortunate that many people think of your contributions and those of
Sam Wilks as coming from the same person!

\textbf{Martin}: Well, the difference between Wilk and Wilks was always
clear to me at least! But you are right that my work with Ram and Sam
led to a host of publications, much of them quite sophisticated
mathematically, too. To an extent, however, this flurry of activity
illustrates a problem I see with mathematical statistics and more
generally with science as I perceive it today. The problem is that there
are a lot of facets to science, and it is now quite fragmented and being
pursued in an opportunistic fashion by individuals in ever more
specialized categories: mathematics, statistics, multivariate analysis,
and so on down the line. I~think this is regrettable.

We need people who look at problems much more broadly, and certainly the
work that Ram Gnanadesikan, Marylin Becker, and I did on the problem of
speech recognition was much broader in nature and of much greater
importance potentially, although our efforts in this direction did not
translate into refereed publications. Ram has given a very good account
of the atmosphere in our research group, and the issues we contended
with, when Jon Kettenring interviewed him for \textit{Statistical
Science} [Vol. 16 (2001), pp. 295--309].

\textbf{Gordon}: It sounds as though you became gradually disenchanted
with mathematical statistics, if not more. Is that why you moved on to
management at AT\&T in 1970?

\textbf{Martin}: I certainly became disenchanted. Through the 1960s, I
acquired a certain sense of the whole organization: I mean Bell Labs, of
course, but also more generally AT\&T. And if I self-examine, I would
say that by 1969, I was waiting for an opportunity to go into
administration.

\section*{Getting Into Management at AT\&T}

\textbf{Christian}: How did that actually happen?

\textbf{Martin}: Through hearings for a rate case that AT\&T had to
submit to in front of the Federal Communications Commission (FCC). At
the time, AT\&T was a huge organization. It was effectively controlling
the local and long-distance telephone network throughout the US, and
because of its quasi-monopolistic position, the FCC had something to say
about its rate of return to equity. One important ingredient in the
hearings held at the end of the 1960s was the so-called Gordon model. It
is a variant of the discounted cash flow model or a method for valuing a
stock or business, if you wish. It was often used at the time to provide
difficult-to-resolve valuation issues for litigation. It was named after
Myron Gordon, who was a finance professor at the University of Toronto.

Now at the time, John Tukey had been asked by AT\&T to criticize some
conclusions that had been derived by the FCC using this model. And John
being the irritating person that he was (because he could figure out
everything so much more quickly than anybody else), I was curious to see
how my hero would fare in the formal and constrained environment of a
federal commission hearing. So I asked permission to attend that part
of the hearings. And at the time, I had no sense of the global issue at
hand, but John talked about this model, and I was a bit amazed to see
that everybody treated him with such great respect. Not that he didn't
deserve it, but litigation tends to be rather merciless.

\textbf{Gordon}: It sure can be. But how did you personally get
involved?

\textbf{Martin}: John's testimony sparked my interest in the Gordon
model, so upon my return to New York, I started to investigate it on my
own, and then I ran quite a few computer simulations. Marilyn Huyett,
who was a close collaborator \cite{8,11,12,13,14,17}, helped me check the
computations. Within about a week, I got a pretty good sense of what was
wrong with the proposed application of this simplistic model. I wrote up
my findings as a technical note and sent it to John Tukey, who suggested
that I pass it on to the upper management at AT\&T. So I did, and then I
got a phone call from Mark Garlinghouse, AT\&T's General Counsel, who
asked me whether I would mind briefing him about the Gordon model.
Amazingly, he had read my report, which was definitely not written from
a business point of view. In fact, part of it was later published in
conference proceedings \cite{35}.

At any rate, Mark Garlinghouse and I met for a whole day shortly
thereafter. We got on extremely well, though we had important political
differences. He listened to me quite carefully and ended up making
representations to the FCC, based on my report. I attended his
presentation and found the experience quite stimulating.

The bottom line is that through that event, the senior people at AT\&T
developed the realization that they needed more attention paid to what
was referred to by them as ``management science.'' Subsequently, I~was invited to spend some time there and help them out with some of their
difficulties. It was a lateral move, shall we say, and at first I did it
as a courtesy to them. But after I had a few interactions, I just got
captured by the effort and started taking an interest in the general
character of AT\&T as an organization.

\section*{Going up the Ranks at AT\&T}

\textbf{Christian}: From 1970 to 1975, you worked at\break AT\&T in various
capacities. You were successively Director of Corporate Modeling
Research (1970), Director of Corporate Research (1971), Director of
Planning (1972), and then Director of Corporate Planning (1973--1975).
What was the bulk of your responsibilities?

\textbf{Martin}: I was initially part of a unit headed by Henry
Boettinger. The function of that small group, less than 20 people in
total, was to look into managerial and financial problems. The group was
actually unique at AT\&T, which had no mathematician, no economist, or
statistician at the time. The organization had basically three
functions: raising money, operating interstate transmissions, and
providing common services for the 22 telephone companies in the AT\&T
group that ran within-state communications. Our group was set up to
formulate plans and policies, to ensure coordination with the developers
at Bell Labs and the implementers at Western Electric.

\begin{figure*}[b]

\includegraphics{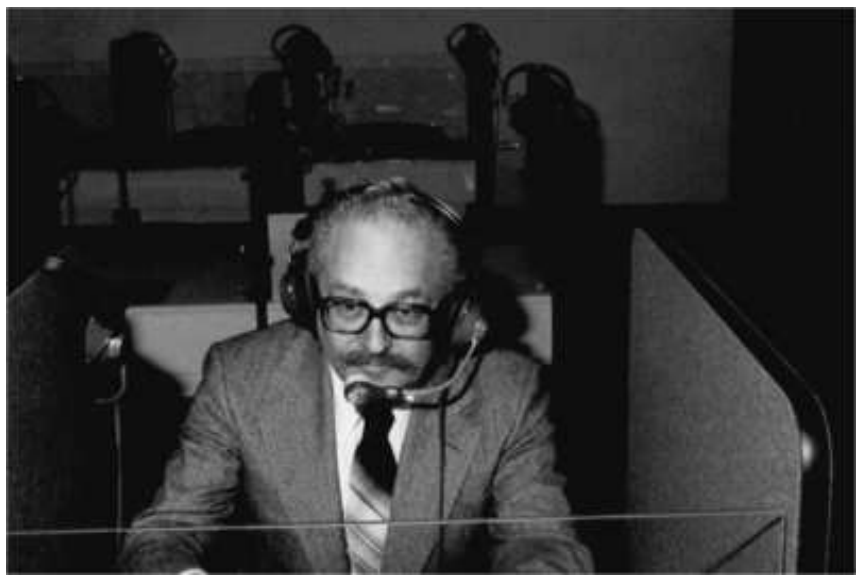}

\caption{Martin brushing up his French soon after joining Statistics
Canada in 1980.}
\end{figure*}

\textbf{Gordon}: With such responsibilities, was it hard for you to
remain active on the statistical front?

\textbf{Martin}: Well, I didn't actually drop technical work completely,
at least in the early 1970s. For instance, I~wrote a paper with Sam
Shapiro on an analysis of variance test for the exponential distribution
\cite{34}. We were subsequently awarded the Jack Youden Prize for the best
expository paper that appeared in \textit{Technometrics} in 1972. I also
managed to get some joint work done with Jane Gentleman \cite{37,38}. But as
time unfolded, the bulk of my responsibilities became more and more
managerial. And of course even more so when I was Assistant
Vice-President and Director of Corporate Planning, from 1976 to 1980.

\textbf{Gordon}: Can you summarize what you accomplished in that period?

\textbf{Martin}: Generally speaking, I carried out work on many large
and small tasks in economics, financial and technological areas,
including cost analysis, registration, and what was then referred to as
``picturephones.''

One major thing that was dumped on me was the Management Research
Information System (MRIS). It had been initially designed to be the
ultimate path-finding operation, a crucial ingredient in operating an
interrelationship between the thousands of pieces and parts that a
telephone system involves.

As it turned out, this MRIS was a colossal mess, and I made it legal to
say so. I found that it was not possible to reorganize it in any sense
or fashion. And the embryonic computing system that was there to make it
alive, supposedly, consisted of parts that could never fit together. In
part, this was because every operating telephone company had a different
computing environment at the time. Also, sadly, some of the people did
not have the expertise and the knowledge required to pull it together.
There just weren't many people around then, who knew how to operate these
systems in a coordinated fashion. Yet the notion of an integrated system
came up every time AT\&T wanted to make an upgrade with what was already
in place! In the end, I got this MRIS closed down.

\section*{Chief Statistician of Canada}

\textbf{Gordon}: How about the transition from AT\&T to Statistics
Canada? This occurred in 1980, right?

\textbf{Martin}: Again, this is not something that I sought. It came up
on its own about five years into my executive position at AT\&T. As it
happens, things were in a bad shape at Statistics Canada at that time,
and the Secretary of the Treasury Board had set up a group under the
headship of Claus Moser to look into the technical competence of the
federal agency. And there was a second, thicker document that had been
prepared by management consultants. At any rate, some member of the
Executive Recruiting Committee phoned me in New York on a Friday
afternoon, either in May or June 1980, to tell me this long story.

\textbf{Gordon}: Sorry. Were you being approached for the position of
Chief Statistician of Canada?

\textbf{Martin}: They probably had that in mind, but the guy was not
explicit about it and it didn't occur to me at first, for two reasons.
First, it was not unusual for senior people at AT\&T to be approached
for counsel, as a service to the public, and given my background, it
would not be completely unexpected for Canadians to seek my opinion in a
period of turmoil at the national statistical agency. Second, I was 57
at the time, so I considered myself too old or too close to retirement
to be perceived as a candidate for the job.

Nevertheless, I was sufficiently interested to look at the reports that
were sent to me after the call, and then to travel to Ottawa to share my
impressions.

\textbf{Gordon}: Who did you meet in Ottawa?

\textbf{Martin}: The meeting was chaired by Jack Manion, who was
Secretary of the Treasury Board of Canada at the time. Other
participants were Harry Rogers, Fred Drummie, and Larry Fry, who was
then interim Chief Statistician. They summarized the two reports
briefly, and then Manion asked me point blank how I would handle the
situation, and why I felt I could do the job! That was a real eye-opener
to me, and I responded by saying that I had come to give my reactions to
the reports, and that I would like to stick to what I came up there to
do. He seemed rather puzzled by my reaction, but we stuck to my agenda.

After I returned to New York, we had an exchange of correspondence and
eventually, I went back to Ottawa for further discussions. The Moser
report was well done, though possibly a little too lenient on the
quality of the surveys and the publications, but it was a fair
assessment. The other report, however, was a terrible document with all
kinds of criticisms that made no sense to me, even though I knew little
about the organization. But what I knew and what I had heard did not
correspond at all to that report's assessment of the situation.

\textbf{Gordon}: So in the end, what convinced you to meet that new
challenge?

\textbf{Martin}: As I reflected upon the subject, I became quite
apprehensive as to what might happen to Statistics Canada if nothing was
done. And the thing that attracted me to take the job is that contrary
to the US, official statistics in Canada was set up as an integrated
system in which data collection, analysis, and so on are coordinated
parts of a grand plan. Anyway, I think I was half decided already when I
was invited for brunch by Michael Pitfield, who was Clerk of the Privy
Council and Secretary to the Cabinet at the time.

\textbf{Gordon}: For the benefit of our readers, it should probably be
said here that the Clerk of the Privy Council is the most senior
nonpolitical official in the Government of Canada. He provides
professional, nonpartisan support to the Prime Minister on all policy
and operational issues that may affect the government.

\textbf{Martin}: Right. And Michael Pitfield, who occupied the position
at the time, was a most obliging person. He gave me a very honest
briefing about the realities of what was going on at Statistics Canada.
He was quite frank, in particular, about the existing fracture lines
within the organization.

\textbf{Gordon}: Before we get to that Martin, could you tell us how
soon after that meeting you accepted the position?

\textbf{Martin}: It took a few more months. You see, my second wife,
Dorothy Barrett, had problems with her eyes at the time. She had to
undergo surgery, and I was not in a good disposition to make a decision
right then.

However, I kept making enquiries about Statistics Canada through the
fall of 1980, while attending to my regular duties at AT\&T. In those
days, there was also a certain amount of reorganization taking place in
the New York head office, and this influenced my hook up.

\textbf{Gordon}: Had you had any real involvement with official
statistics prior to becoming Chief Statistician of Canada?

\textbf{Martin}: I was on the US Census Advisory Committee in the early
1970s. I remember urging them to develop some sort of strategic plan for
evolution and commit it to paper. And I remember making an argument
(which I subsequently made at AT\&T too) that although the plan may not
be fully implemented, it would at least force them to think where they
were, where they wanted to go, and how to get there. Also, it is
important to put a calendar to it. At a later stage, I~remember making a
number of concrete suggestions to the US Census Director, Vince Barabba,
and they were well received.

\section*{Main Challenges at Statistics Canada}

\textbf{Christian}: Can you summarize the issues that you faced when you
arrived at Statistics Canada?

\begin{figure}

\includegraphics{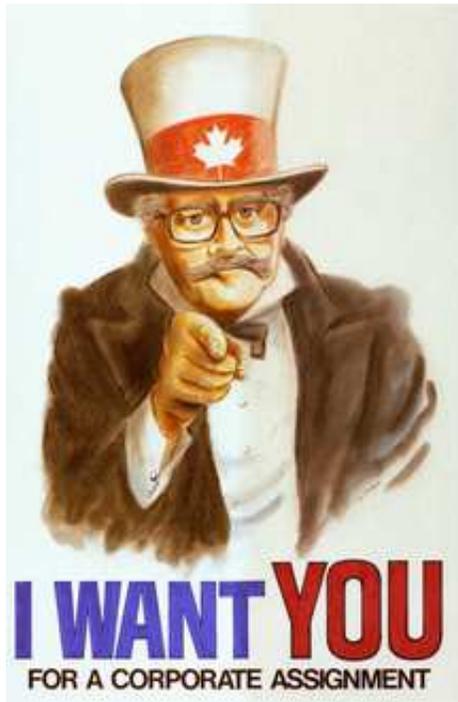}

\caption{``Uncle Martin'' in a poster designed by Statistics Canada and
adopted widely in the Canadian Government to publicize a program
launched in 1984--1985 to enable civil servants to change assignments.}
\end{figure}

\textbf{Martin}: One major organizational problem was the division of
social statistics and economic statistics into two almost completely
separate units. These sections were headed by Ivan Fellegi and Guy
Leclerc, respectively. Each of them had a title of Assistant Chief
Statistician. The survey methodology people were also split between
these two sections.

In addition, there was a major problem in human relations and in the
image that the organization was projecting. It affected its credibility
in the eyes of the public. There were allegations of nepotism in the
press and many good people ended up leaving the organization because of
the extremely negative climate that prevailed. Jacob Ryten was one of
them. And you left too, Gordon, isn't that right?

\textbf{Gordon}: Yes. I went to work for the British Colum\-bia
statistical agency for a while.

\textbf{Martin}: The morale at Statistics Canada was very low, and this
was at a time when the 1981 Census had to be carried out. Mind you, the
planning for that operation was well underway. At the heart of it was Ed
Pryor. It was a major undertaking, but we were globally satisfied with
it.

\textbf{Gordon}: Luckily, by the end of your term, staff relations had
improved considerably.

\textbf{Martin}: I am glad it worked out okay. I certainly tried to
assist every way I could. One person who helped me a great deal in
sorting out the staff problems was Jean-Jacques Blais, who was the
Minister responsible for Statistics Canada when I took office. We
developed a very good rapport. In addition, the Treasury Board was
prepared to give me some slack, and that helped me considerably in
dealing with some individuals. There was a fair amount to be done in
this regard, but there is no point talking about this, as it would
involve people by name and so forth. Suffice it to say that there was a
certain amount of unpleasantness about it, but not much beyond that. And
ultimately, staff movements helped resolve the separation between social
and economic statistics.

\section*{Other Issues}

\textbf{Christian}: Besides human relations, could you give us an idea
of the issues you had to deal with?

\textbf{Martin}: There were many on the methodological side. And to be
truthful, I didn't have much of a clue at first as to what was going on
in terms of production, i.e., what people did, how they did it, and why
they did it.

One major problem for me was that I couldn't afford to take the time to
find out in depth, because I needed speed of reaction that was far
greater than my speed of learning. For example, there were major
discrepancies in employment figures at the time between the Labour Force
Survey and the Survey of Employment Payroll and Hours. The problem was
of considerable importance and urgency because these particular surveys
attract a lot of attention in the public.

Another major issue that troubled me very early on was the range of
publications that was turned out by Statistics Canada. There seemed to
be no rhyme or reason to much of it. Much time, effort and money was
being wasted in producing publications that very few people would read.
The publication issue was eventually resolved; many significant changes
were made. It was a long battle, and many compromises needed to be
struck in order to make publications financially self-sufficient.

\textbf{Gordon}: These efforts resulted in fewer publications, and more
importance to those retained.

\textbf{Martin}: Right. On another front, I promoted a form of
operational integration within headquarters in Ottawa, and also
regionalization of the data collection operations. I tried to give more
importance to the regional offices in Halifax, Montr\'{e}al, Toronto,
Winnipeg, and so on. The process has expanded since, but setting it up
involved a lot of fighting initially. I ended up having to move
a lot of people around, but in so doing I always tried to avoid
disrupting their career goals.

And yet another important battle I recall was getting access to cabinet
documents. There were strategic reasons behind this action, but it was
part of a concerted effort to bring Statistics Canada into a closer
relationship with other departments. This is one thing that struck me
when I first joined the agency: it seemed that Statistics Canada
operated almost entirely on its own, and that it had very few contacts
with other branches of government. I needed to make the personal effort
of contacting the Deputy Ministers in charge of the various departments
to set up working committees that could oversee the information and data
collection needs of these folks, and establish mechanisms to meet their
demands.

\begin{figure*}[b]

\includegraphics{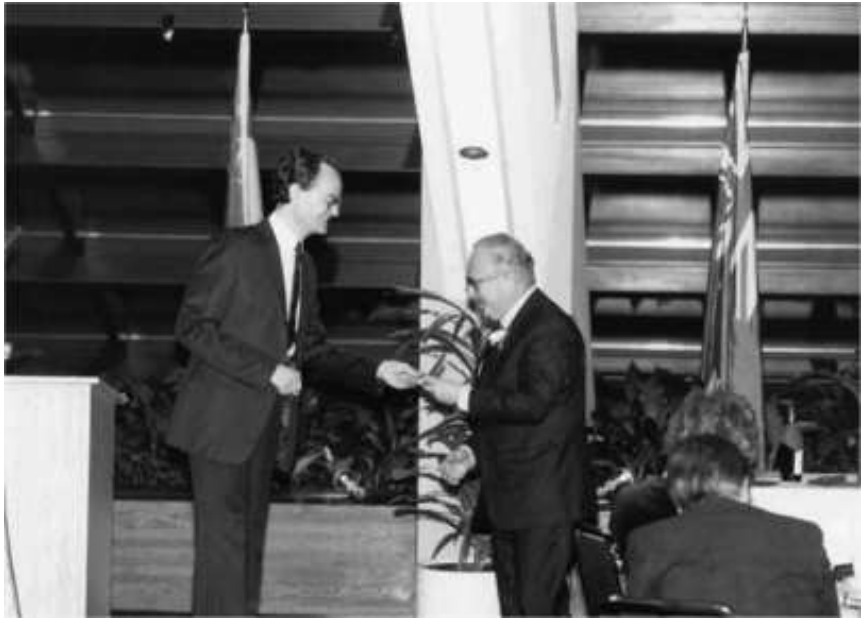}

\caption{Martin Wilk at his retirement party, January 1986, receiving
from his successor, Ivan Fellegi, the keys of a motor scooter, a gift
from the employees of Statistics Canada.}
\end{figure*}

\textbf{Christian}: By setting up external advisory groups, you also
tried to help the agency in its efforts to improve its effectiveness and
coverage and it ultimately led, in the early 1990s, to international
marks of recognition for the world-class quality of its statistics.

\textbf{Martin}: Yes, that was the long-term plan, and it was satisfying
to see it materialize. To this end, I established quite a few of these
advisory committees, maybe a dozen. And many of them have survived to
this day. In fact, were you not a member of the Advisory Committee on
Statistical Methods with me in the late 1990s?

\textbf{Christian}: That's right. I was on the committee from 1994 to
1999, I think.

\textbf{Martin}: I should take this opportunity to say that the most
prominent of all these committees was actually set up by my successor,
Ivan Fellegi.

\textbf{Gordon}: You mean the National Statistics Council, right?

\textbf{Martin}: Absolutely.

\section*{The 1986 Census}

\textbf{Gordon}: Maybe we should move on to the 1986 Census now? I know
you had just retired when it was run but as I recall, it was nonetheless
a major issue for you.

\textbf{Martin}: Oh yes, indeed. When Brian Mulroney became Prime
Minister of Canada, in September 1984, he decided to make a clean sweep
and undertook a number of initiatives. As you may remember, his party
won the largest majority government in Canadian history. One of
Mulroney's priorities was to control the deficit, which was running into
the billions of dollars. He started seeking contributions from every
department that would result in monetary savings. I was not initially
consulted on this issue but it turned out that the Minister responsible
for Statistics Canada volunteered on his own to cancel the 1986 Census!

\textbf{Christian}: Who was this?

\textbf{Martin}: It was Harvie Andre, a pleasant enough fellow from
Alberta who, incidentally, had a Ph.D. in Chemical Engineering. But he
was quite opinionated and didn't have a high regard for statisticians.
He could see substantial savings in cancelling the 1986 Census, and so
he had simply gone ahead with it!

When I first heard about this, it was underway. When I found out, I
really had a rush of anger. I felt he had absolutely no right to make
such a judgment without consulting me at least. As I recall, I really
blew him.

\textbf{Gordon}: What happened next?

\textbf{Martin}: Given the circumstances, my first responsibility was to
establish a high-level committee internally to review what would need to
be done to meet the requirement if it came to it. Second, in parallel to
that, I encouraged people from outside the agency to make their feelings
known as to how much their operations would be affected if the census
were cancelled.

\begin{figure*}[b]

\includegraphics{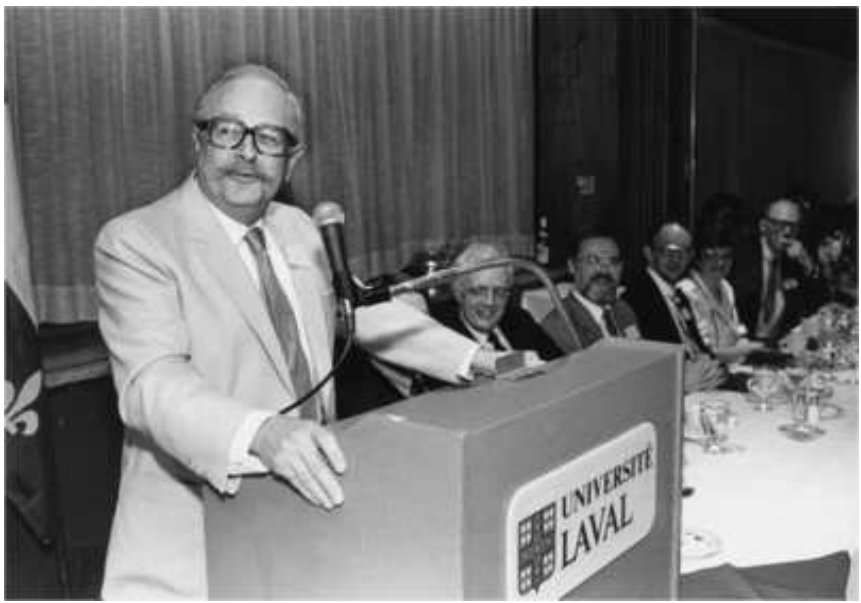}

\caption{Martin addressing SSC members at the banquet held during the
1987 SSC Annual Meeting at Universit\'{e} Laval, Qu\'{e}bec. Guests at
the table of honor in the background (from left to right): Sir David
Cox, Terry Smith, Marc Moore and his wife, and Michael Stephens.}
\end{figure*}

Pretty soon, we found out that by law, the Prairie provinces had to have
a census in 1986. And also, that the lack of census would affect
seriously a host of programs within and outside the agency. Harvie Andre
and his staff came to recognize their mistake, and so the census was
reinstated. But in the meantime, the commitment had been made that
Statistics Canada would cut off 100 million dollars from its budget.
That was an awful lot of money; maybe a third of the agency's entire
budget!

\textbf{Christian}: So it looks like the politician reached his goal
anyway. How did you get around that one?

\textbf{Martin}: Through intense rounds of discussions with government
representatives like Jack Manion and Harry Rogers, it was agreed that
the money we had already saved by integrating our operations and
rationalizing our publications would be included in the 100~million. At
the end of the day, that left me looking for some 40 million only!

\textbf{Gordon}: That's still a hefty sum of money.

\textbf{Martin}: Through additional negotiations and reflection, we came
up with a big package in which Statistics Canada absorbed maybe 10
million dollars on its own. Another chunk came through contributions
from some departments that had vested interests in the census. And then
the rest came as a credit for student hirings. You see, we estimated
that over 40,000 people would have to be hired to run the census, but
the money was not actually needed until 1986!

\textbf{Gordon}: All this happened within a very short period of time
too. It must have been quite a stress.

\textbf{Martin}: It sure was, and I even got physically ill over it. So
once the 1986 Census was officially reinstated, I~thought it was a
reasonable time for me to leave. I~mean it was a big victory, and at the
time I felt there was little more that I could do, at least in the short
run, to establish Statistics Canada on an on-going basis. Bear in mind
also that I was 63 at the time.

Now after Guy Leclerc transferred to the Secretariat of the Treasury
Board, back in 1983, I had arranged for Ivan Fellegi to be promoted to a
newly created position of Deputy Chief Statistician. I chose that
designation to make it clear that he was a natural candidate for my
succession, and when the time came for me to retire, I~pretty well made
it a condition that Ivan would be my replacement.

As Deputy Chief Statistician, he had begun to think of the organization
as a whole, including economics, and he had done very well, making fast
progress. So he was pretty well the natural candidate, and Prime
Minister Mulroney followed the recommendation made by Privy Council.
This way, I could retire in September 1985, and Ivan had enough time to
get established before the 1986 Census was run. He has done very well
since and is now, I believe, in his twentieth year as Chief Statistician
of Canada. [Note: Dr Fellegi retired from that position on June 12,
2008.]

\begin{figure*}[b]

\includegraphics{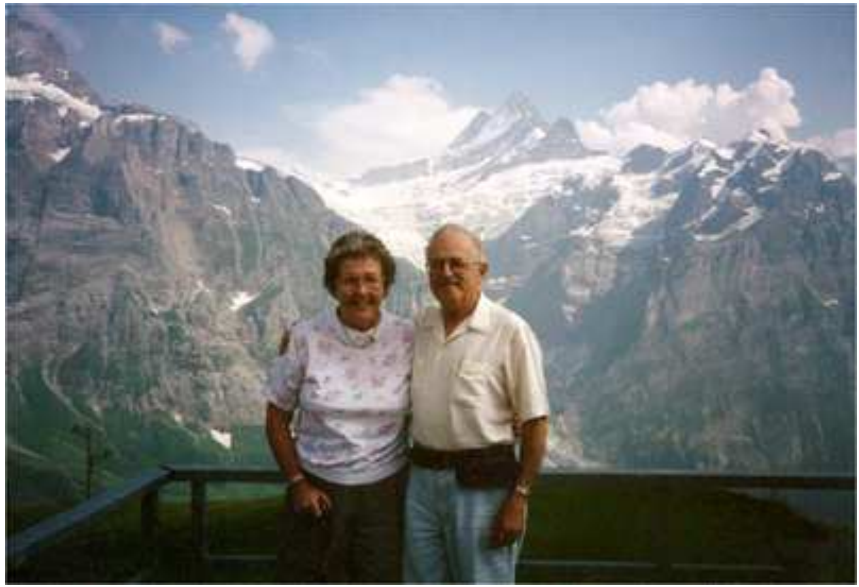}

\caption{Martin Wilk and his second wife, Dorothy Barrett, at the
Schreckhorn peek in Grindelwald (Switzerland), circa 1993.}
\end{figure*}

\section*{Presidency of the SSC}

\textbf{Christian}: Martin, when you had your official retirement party
in January 1986, you were just starting your term as President of the
Statistical Society of Canada (SSC).You had already made quite an
impression on its members at the 1985 Annual Meeting in Winnipeg. Many
of us remember vividly your after-dinner speech on ``blue-collar and
white-collar statisticians.'' There was a lot of wisdom in that talk
which, luckily, was later published in \textit{Survey Methodology} \cite{42}.

\textbf{Martin}: In fact, it is even available in French \cite{43} because
this journal, also known as \textit{Techniques\break d'enqu\^{e}te}, is
published in both languages by Statistics Canada.

\textbf{Christian}: Right. And because of a change in the SSC By-Laws,
you were in fact the only person to serve as President for more than a
year. Could you tell us about your term?

\textbf{Martin}: When I was voted in, I found that the SSC was a pretty
messy organization. It was still suffering, I think, from the
unfortunate circumstances that led to its creation, in 1978. Of course,
you already know that story, Christian, having related it as you did
with David Bellhouse in some past issue of \textit{Statistical Science}
[Vol. 14 (1999), pp. 80--125].

Anyway, I had an orientation at the time of making statistics better
known and more visible in\break Canada. One way of accomplishing that was to
elevate the level of the \textit{SSC Newsletter} to a full-fledged
publication. I was fortunate that Nicole Gendreau agreed to take on that
mandate, and to involve the resources of the Bureau de la statistique du
Qu\'{e}bec that she was head of at the time. That's how SSC's
\textit{Liaison} emerged as the quarterly, fully bilingual magazine that
it continues to be to this day.

Another major effort was the in-depth revision of the SSC By-Laws, which
Peter Macdonald and you, Christian, put a lot of effort into.

\textbf{Christian}: When these By-Laws were approved by the Board of
Directors, I remember you congratulating me on this effort and telling
me ``And now, young man, on to greater things!'' I was 30 then. It made
a big impression on me. I've been trying to live up to the challenge
since!

\textbf{Martin}: These volunteer jobs are essential for the profession,
but they can be really time-consuming. I always tried to do my share,
having served as President of my local chapter of the American
Statistical Association, and then as Vice-President at the national
level, from 1980 to 1982. I was also a member of the founding Editorial
Board for \textit{Technometrics} (1959--1963).

\section*{After Retirement}

\textbf{Gordon}: You certainly had a very full professional life, and
your achievements have been underscored with many honors. From what I
know, you didn't really slow down after retirement. Could you summarize
very briefly?

\textbf{Martin}: At first, I served as Senior Advisor to the Privy
Council Office, say for a period of six months. Afterward, I did quite a
bit of consulting for Statistics Canada, Revenue Canada, the US Bureau
of the Census, the Ontario Premier's Council, and so on. I also served
as a member, and sometimes as a Chair, on various bodies both in Canada
and in the USA, where my wife and I chose to go back in recent years.

One of my major assignments was on the Canadian Institute for Advanced
Research, whose mission is to orient and promote research in Canada. The
Institute might identify nanometrics, say, as an area of opportunity for
the country, and then there would be a task force set up to explore that
possibility, and so on. The~Population Health Program was one of the
most visible initiatives launched by this institute.

\textbf{Christian}: But your involvement with the Task Force on Health
Information was even greater. Right?

\textbf{Martin}: Indeed! At the time there was a public sense that
statistical information about the performance of Canada's health system
was either nonexistent or fragmented, health being a matter of
provincial jurisdiction. The Task Force report, that I wrote much of
myself, was well received and led to a revamping of the organization and
management of health statistics. This included, for example, the
creation of the Canadian Institute for Health Information to complement
the survey work of Statistics Canada by gathering and analyzing
operational information from health institutions across the land.

One other important committee I got involved with had to do with
science. Curiously, Statistics Canada had never gotten really interested
in the area of science. Our objective, or at least one of them, was to
report on the characteristics and character of science as being pursued
in Canada. Members of the steering committee included Jacob Ryten,
Steven Fienberg, as well as Mike Sheridan,\break Michael Wolfson, Scott
Murray, Ray Ryan, and possibly others.

\textbf{Christian}: And for quite a while, of course, you remained
active with Statistics Canada as a member of its Advisory Committee on
Statistical Methods!

\textbf{Martin}: I enjoyed that a lot. But I am now over 80 and I have
pretty well given up on all these things. These days, I just try to
enjoy myself with my wife Dorothy, my children and my grand children.

\textbf{Christian}: You certainly deserve it! Is there anything you
would like to add before we close?

\textbf{Martin}: With your permission, there are several persons that I
would like to thank here because of the important roles that they played
in my career. First and foremost, I would like to mention my wife,
Dorothy and my children, Rebecca, Carol, David, Teresa, and Kathryn, who
saw me through both difficult and very pleasant times. Then from a
professional point of view, I was also quite fortunate to find many
collaborators and friends. I am especially grateful to Oscar Kempthorne,
John Tukey, Ram Gnanadesikan, Henry Boettinger, Paul Reed, Harry Rogers,
Ivan Fellegi, Jacob Ryten, and Fraser Mustard. Needless to say, the list
is not exhaustive.

\textbf{Gordon}: Thank you very much, Martin, for all the time you have
given us.

\textbf{Martin}: And thanks to you guys for traveling to Oregon and for
the time you devoted to me.

\section*{Acknowledgments}

Thanks are due to Marianne Genest for transcribing the interview.
Funding in partial support of this work was provided by the Natural
Sciences and Engineering Research Council of Canada, the Fonds
qu\'{e}b\'{e}cois de la recherche sur la nature et les technologies, and
the Institut de finance math\'{e}matique de Montr\'{e}al.

\vspace*{-2pt}

\end{document}